\documentstyle[preprint,aps,prb,epsfig]{revtex}
\begin{document}
\draft
%\twocolumn[\hsize\textwidth\columnwidth\hsize\csname
%@twocolumnfalse\endcsname
\title{
An electron paramagnetic resonance study of  Pr$_{0.6}$Ca$_{0.4}$MnO$_{3}$
across the charge ordering transition
}
\author{ Rajeev Gupta, Janhavi P. Joshi, S. V. Bhat and A. K. Sood}
\address{
Department of Physics,
Indian Institute of Science,
Bangalore 560 012, India
}
\author{C. N. R. Rao}
\address{
CSIR Centre for Excellence in Chemistry, Jawaharlal Nehru Centre for Advanced
Scientific Research, Jakkur P.O, Bangalore 560 064, India}
\maketitle
\date{today}
\widetext
\begin{abstract}
{We report the first electron paramagnetic resonance  studies of  single
crystals and powders of Pr$_{0.6}$Ca$_{0.4}$MnO$_{3}$  in the
300-4.2 K range, covering the charge
ordering transition at $\sim$ 240 K and antiferromagnetic transition (T$_N$)
at $\sim$ 170 K. The asymmetry parameter for the Dysonian single crystal
spectra shows anomalous increase at T$_{co}$. Below T$_{co}$ the g-value
increases continuously, suggesting a gradual strengthening of orbital
ordering. The linewidth undergoes a sudden increase at T$_{co}$ and continues
to increase down to T$_N$. The intensity increases as the temperature is
decreased till T$_{co}$ due to the renormalization of magnetic susceptibility
arising from the  build up of ferromagnetic correlations. The value of
the exchange constant, $J$,  is estimated to be 154 K. }

\noindent{PACS numbers: 76.30.-v, 75.70.Pa, 72.80.Ga, 71.30.+h \\ }
\end{abstract}
%]

%\narrowtext

%\section{ Introduction}
Recent investigations of rare earth  manganites and related
systems exhibiting colossal magnetoresistance
has led to the discovery of interesting phenomena related to charge, spin
and orbital ordering in these materials \cite{cnr2}.  Phase diagrams
of these systems as a function of doping and temperature are therefore very
interesting showing regimes of varied magnetic and electrical properties.
These manganites of general composition
$A_{1-x}A^{\prime}_{x}MnO_{3}$ where $A$ is a trivalent rare earth ion (e.g
La, Pr, Nd etc) and $A^{\prime}$ is a divalent ion (e.g Ca, Sr, Pb etc) show a
rich phase diagram \cite{cnr1} depending on the tolerance factor and the
amount of doping x which, in turn, controls the ratio of  Mn$^{3+}$
to Mn$^{4+}$. The end members of this series i.e x=0 and x=1 are A type and G
type antiferromagnetic insulators (AFI), respectively. For 0.17 $<$ x $<$ 0.5
the system undergoes a metal to insulator (MI) transition as the temperature
is increased. This MI transition also coincides with the magnetic transition
from the ferro to the paramagnetic state.  Electronic
properties of these systems are understood qualitatively in terms of the
Zener's double exchange interaction \cite{zene,ande,genn} (DEX).
In DEX there is a strong Hund's coupling
between the 3 $t_{2g}$ electrons whose spins are parallel and constitute a
core spin of 3/2 and the lone $e_g$ spin. This $e_g$ electron can hop from one
Mn$^{3+}$ site to adjacent Mn$^{4+}$ site via the intermediate oxygen when the
core spins are parallel which also implies that metallicity coincides with
ferromagnetism. To explain the resistivity data, Millis et al
\cite{mill} invoked the localization of charge carriers above T$_c$ due to
polaron formation. In the case of systems where the weighted
average A site cation radius is small, the system also shows charge ordering
(CO) i.e real space ordering of Mn$^{3+}$ and Mn$^{4+}$ ions, as a function of
temperature. The CO state becomes stable when the repulsive Coulomb
interaction between carriers is dominant over the kinetic energy.
In these types of cases there is a strong competition between the double
exchange interaction which favours ferromagetism and CO which favours
antiferromagnetism. The smaller size of A site cation leads to a deviation in
the Mn-O-Mn bond angle from 180$^{\circ}$ resulting in lowering of the
transfer integral. This in turn implies a lower band width of the $e_g$
electron and hence  higher electronic correlations in the system. The
stability of the CO state depends on the commensurability of the carrier
concentration with the periodicity of the crystal lattice  and is stable for
x=0.5.

Pr$_{1-x}$Ca$_{x}$MnO$_{3}$ shows a rich phase diagram as a function of
doping and temperature and is well studied using a variety of probes like
resistivity \cite{tom1}, magnetization \cite{lees}, neutron diffraction
studies \cite{yosh,cox1,jir1,jir2} and transmission electron microscopy
\cite{mori}. For x=0.4, the sample is insulating at all temperatures in zero
field. The resistivity changes by more than six orders of magnitude from 300 K
to 50 K. There is a perceptible change in slope at around T$_{CO}$ $\sim$ 240
K signifying the onset of charge ordering.  In the temperature range T $>$
T$_{co}$ (240 K) the system is a paramagnetic insulator. In dc magnetic
susceptibility, a large peak is observed at T$_{co}$ followed by a relatively
small peak at T$_{N}$. The peak at T$_{co}$ is attributed to ferromagnetic
correlations \cite{cox1}. The system further undergoes a transition at 170 K
to a CE type AFI. The pseudo CE type structure for x $<$ 0.5 is different than
the CE structure  present in systems with x = 0.5. In this so called "pseudo
CE" structure the zig-zag FM chains in the ab plane are FM aligned along the c
axis \cite{cox1}. This is unlike the CE structure where the layers in the ab
plane are aligned antiferromagnetically along c. Further lowering of
temperature below 50 K leads to another transition which can be understood
either as a canted antiferromagnetic state \cite{jir1,jir2} or as a mixture of
ferromagnetic domains or clusters in an antiferromagnetic background.
In this paper we report our EPR study of single crystals as well as powders of
Pr$_{0.6}$Ca$_{0.4}$MnO$_{3}$  as a function of
temperature from 300 K to 4.2 K. Although several  EPR studies  on the
manganites have been reported recently  across the metal-insulator transition
\cite{hube,osef,riva,caus,tova,shen,lofl,ivan}, to our knowledge, there has
been  no study  across the CO transition. The
dynamics of spins in the charge ordered state as studied by EPR is expected to
throw some light on the controversial magnetic structure between the charge
ordering transition and the Neel temperature.

%\section{Experimental details}

The single crystal used for
the experiment was prepared by the float zone technique
and characterized using dc magnetic susceptibility which shows a large peak
at 240 K (T$_{co}$) and a relatively small peak at 170 K (T$_N$). The
measurements on single crystals were done with magnetic field parallel to
(100) axis. Powder of the material was dispersed in paraffin wax for study.
The EPR measurements were carried out at 9.2 GHz (X band) with a Bruker
spectrometer ( model 200 D) equipped with an Oxford Instruments continous flow
cryostat ( model ESR 900), with a temperature accuracy of $\pm$ 2 K.

%\section{Results and Discussion}
 Fig. 1 shows the EPR spectra at a few temperatures in the
range 300 K to 180 K recorded in the heating run for a single crystal
(fig. 1(a)) and a powder sample (fig. 1(b)). The ESR signal could be observed
only above 180 K. The observed signals are very broad.
The spectra from the single
crystal are Dysonian in shape and were fitted (solid lines in Fig. 1(a)) to a
functional form similar to the one used by Ivanshin et al \cite{ivan} for
La$_{1-x}$Sr$_{x}$MnO$_3$. The field derivative of the power absorbed given by

\begin{equation}
\frac{dP}{dH} \propto \frac{d}{dH}(
\frac{ \Delta H + \alpha (H -
H_{res})}{(H - H_{res})^{2} + \Delta H^{2}} +
\frac{ \Delta H + \alpha (H +
H_{res})}{(H + H_{res})^{2} + \Delta H^{2}})
\end{equation}
incorporates responses to  both the circular components of
the exciting linearly polarised microwave field. The above equation
also includes both absorption and dispersion. $\alpha$, the asymmetry
parameter is a measure of dispersion-to-absorption ratio. The spectra from
powder samples are symmetrical and were well fitted with Lorentzians as shown
in Fig. 1(b).

Careful  EPR
measurements in doped manganites by Causa et al \cite{caus} and Lofland et al
\cite{lofl} show that that both Mn$^{3+}$ and Mn$^{4+}$ contribute to the EPR
signal. The bottleneck model used by Shengalaya et al \cite{shen} also shows
that the EPR intensity is proportional to the total susceptibility of the
Mn$^{4+}$ and Mn$^{3+}$ spins. We will first discuss our results on
temperature dependence of EPR lineshape parameters of single crystals.

Fig. 2 shows the temperature dependence of the lineshape parameters of the
signals obtained by fitting to equation (1).  Fig. 2(a) shows the temperature dependence of the
linewidth ( $\Delta$ H). The linewidth decreases as the temperature is lowered
from 300 K upto just above T$_{co}$ and then nearly doubles across the charge
ordering transition. This increase is somewhat similar to that observed
\cite{shen} in LaCaMnO$_3$ across the paraamgnetic insulator to
ferromagnetic metal transition, thereby implying a buildup of
spin-correlations at T$_{co}$. The increase in linewidth above T$_{co}$ can be
interpreted in terms of spin-lattice relaxation or even as an opening up of
the ``bottleneck'' as the temperature is raised \cite{shen,lofl}. Fig. 2(c)
shows the variation of the integrated intensity as a function of temperature.
We have tried to analyse the temperature dependence of intensity in terms of
the bottleneck model \cite{shen,koch}. In this picture, the EPR signal
originates from both Mn$^{4+}$ and Mn$^{3+}$ ions and the intensity is
proportional to the total susceptibility.
Here the spin-spin relaxation rates
between the exchange coupled Mn$^{3+}$ and Mn$^{4+}$ ions are much larger than
the spin-lattice relaxation rates. In this regime the ferromagnetic
correlations will renormalize the spin susceptibility given by \cite{shen}
\begin{equation}
I \propto \chi_{total} = \chi_{s} + \chi_{\sigma}
\end{equation}
where $\chi_{s}$ and $\chi_{\sigma}$ are the renormalised static
susceptibilities and are given by
\begin{equation}
\chi_{s} = \chi^{o}_{s} \frac{1+
\lambda \chi^{o}_{\sigma}}{1- {\lambda}^2 \chi^{o}_{\sigma} \chi^{o}_{s}} ,
\hskip30pt  \chi_{\sigma} = \chi^{o}_{\sigma} \frac{1+
\lambda \chi^{o}_{s}}{1- {\lambda}^2 \chi^{o}_{\sigma} \chi^{o}_{s}}
\end{equation}
where $\chi_{s}^{o}$ = bare spin susceptibility of Mn$^{4+}$
and $\chi_{\sigma}^{o}$ = bare spin susceptibilty of Mn$^{3+}$. The parameter
$\lambda = zJ/Ng_{s}g_{\sigma} \mu_{B}^{2}$ , where $J$ is the exchange
coupling constant between Mn$^{4+}$ and Mn$^{3+}$ spins, $g_{s}$
($g_{\sigma}$) is the g factor of Mn$^{4+}$ (Mn$^{3+}$) ions ,$N$ is the
number of spins per cm$^{3}$,  $z$ number of nearest neighbours and $\mu_B$ is
the Bohr magneton.
We can get some estimate of $J$ by fitting Eqs. 2 and 3 to the data in Fig
2(c). Taking $g_{s}$ = $g_{\sigma}$ = 2, $z$ = 6 and assuming that the bare
susceptibility of Mn$^{4+}$ ions is given by Curie law $\chi^{o}_{s}$ =
C$_{s}$/T while that of Mn$^{3+}$ follows a Curie-Weiss law
C$_{\sigma}$/(T-$\Theta$), where $\Theta$, the negative Curie Weiss
temperature is taken to be the same as that in undoped LaMnO$_3$ ($\Theta$ =
-100 K). The solid line shows the fit to the intensity data above T$_{co}$
yielding an estimate of $J \sim$ 154 $\pm$ 1.24 K.  This value is of the same
order of magnitude as  the value obtained for $J$ ( $\sim$ 70 K) in
doped manganites by EPR measurements \cite{shen}, neutron scattering
\cite{dai} and Brillioun scattering experiments \cite{bhas}. At this stage we
would also like to point out that though all spins contribute to the EPR
intensity, the temperature dependence is qualitatively different from the dc
susceptibility data (as shown in inset of Fig2(c) ) . This could be due the
fact that the resistivity of the sample is increases by two orders of
magnitude in the temperature regime of 300-180 K which will enhance
the penetration depth and hence the volume of sample seen by the microwave
field is changing with temperature.
Fig 2(b) shows the temperature variation of g value. Since internal field
effects can influence the single crystal data, we will focus our attention on
temperature variation of g only in the powder data.

 'Asymmetry parameter',
 $\alpha$  of the signals, from single crystal data is shown
in Fig. 3. The ratio remains practically constant from 300 K to
T$_{co}$ where it discontinously increases from $\sim$ 2.5 to $\sim$ 4.5.
Further cooling results in a decrease of $\alpha$ value towards unity
as is to be expected from the decrease in conductivity at lower temperatures
(resistivity \cite{ayan} shown in inset of Fig. 3). The discontinous {\em
increase} of $\alpha$ at T$_{co}$ is interesting because normally one would
have expected a decrease following the decrease in conductivity. However as
shown by Dyson \cite{dyso} and Feher and Kip \cite{fehe} the value of $\alpha$
depends also on the ratio of the time $T_D$ taken by the electrons to diffuse
through the skin depth and the spin-spin relaxation time $T_2$. The sudden
increase in $\alpha$ therefore indicates a sudden dip in the value
$T_D$/$T_2$. This implies that the value of $T_2$ increases at T$_{co}$ to
offset the increase in  $T_D$ due to decrease in the conductivity. However,
this is contrary to the observed increase in the linewidth at T$_{co}$. At
present, we do not have a satisfactory explaination of this result.

 Fig 4 shows the temperature dependence of the lineshape parameters for
the powder sample extracted by fitting derivative of the Lorentzian function
yielding temperature dependence of the linewidth ($\Delta
H$), resonance field ($H_{o}$ ) and area under the curve.
 Fig. 4(b) shows the temperature dependence of the g factor, estimated from
the resonance field. In all the earlier EPR reports in manganites, the value
of g was observed to be close to or less than that of free electron ( =
2.0023). Earlier EPR studies on Mn$^{3+}$ and Mn$^{4+}$ dilutely doped in
diamagnetic hosts have given \cite{pilb}  g $\sim$ 1.98. However our
experiments give a g value  higher than that for the free electrons for all
temperatures. A speck of DPPH ( g = 2.0036) was used as a g-marker and the
centre resonance field was obtained from the fit to the Lorentzian. This
procedure gives confidence in our measurements of temperature dependence of g.
 Since the internal field effects are expected to average out in powdered
samples, we believe that the increase in g for T $<$ T$_{co}$ is intrinsic in
nature. One possible reason can be the changes in the spin-orbit coupling.
It is known that  g$_{eff}$ = g[1
$\pm$ $\kappa$/ $\Delta$], where the spin orbit interaction energy is $\kappa
\vec{L}. \vec{S}$ and $\Delta$ is the appropriate crystal field splitting
\cite{kitt}. A recent transmission electron microscopy study \cite{mori} of
Pr$_{0.5}$Ca$_{0.5}$MnO$_3$ has shown that the incommensurate to commensurate
charge ordering is coincident with paramagnetic to antiferromagnetic
transition at 180 K. The physical picture is that for T$_{N}$ $<$
T $<$ T$_{co}$, orbital ordering is partial inspite of complete charge
ordering. This orbital ordering builds up and is complete at T$_N$. Therefore
for T $<$ T$_{co}$ a gradual buildup of the orbital ordering can change the
spin-orbit coupling and lead to an increase in the value of g as the
temperature is lowered. This may also be the reason for the sign of g shift
with respect to the free electron g which is opposite to that expected for
Mn$^{3+}$ and Mn$^{4+}$ ions ( less than half filled d-shells). It will be
very interesting to theoretically calculate the value of g, incorporating
orbital ordering. Fig. 4(a) shows the temperature variation of the full width
at half maximum of the Lorentzian, $\Delta H$. For T $>$ T$_{co}$, the width
decreases linearly as the temperature is lowered from 300 to 240 K similar to
that observed in single crystal data.  Below T$_{co}$ till 180 K, the width
increases significantly (almost by a factor of 2). It is remarkable that the
width increases sharply from $\sim$ 1600 G at 240 K to $\sim$ 2200 G at 228 K.
This sharp increase can be due to magnetic fluctuations which are also
responsible for the peak in the dc magnetic susceptibility. The further
increase in $\Delta H$ as temperature is lowered can arise due to build up of
magnetic correlations preceeding the transition to the long range
antiferromagnetic ordering at 170 K. Fig. 4(c) shows the temperature
dependence of the intensity which is qualitatively similar to that of single
crystal data.

% \section{Summary}
In summary, we have reported for the first time EPR measurements on
charge-ordered  Pr$_{0.6}$Ca$_{0.4}$MnO$_3$. The lineshape parameters reveal a
rich temperature dependence across T$_{co}$ as well as at  lower
temperatures approaching T$_N$. The intensity variation above T$_{co}$ is
considered to be due to the renormalization of spin
susceptibility due to ferromagnetic (FM) correlations. Such FM correlations in
the paramagnetic insulating phase have been invoked to understand the origin
of the peak in magnetic susceptibility near T$_{co}$ and of dynamical spin
fluctuations above T$_c$ in doped manganites \cite{lynn,tere}. The formation
of magnetic polarons to localize the carriers has been suggested by Varma et
al \cite{varm}. The value of  exchange coupling constant $J$ is estimated
to be about 154 K. The fluctuations in the magnetic correlations near the
transition temperatures lead to a large increase in the linewidth.
The temperature dependence of the g factor  suggests a need
to carry out theoretical calculations of g invoking orbital ordering.

SVB and AKS thank the Department of Science and Technology for financial
assistance.  We thank Prof. B. S. Shastry for his suggestion to do the EPR
experiments and for useful discussions.

\begin{figure}
%\centerline{\psfig{figure=finepr1.eps,width=15cm,height=17cm}}
\label{Fig 1}
\caption{Derivative spectra of Pr$_{0.6}$Ca$_{0.4}$MnO$_{3}$
for (a) single crystal and (b) powder sample at a few select temperatures.
The signal from DPPH has been subtracted from fig. 1(b).
The solid lines  shows the Dysonian and Lorentzian fits
to the crystal and powder sample data respectively.}
 \end{figure}

\begin{figure}
%\centerline{\psfig{figure=finepr2.eps,width=15cm,height=17cm}}
\label{Fig 2}
\caption{ Variation of the lineshape parameters - linewidth, g and
intensity as a function of temperature for the single crystal data obtained by
fitting to eq. (1) . The solid line in Fig. 2(c) shows the fit for T $>$
T$_{co}$ to Eq. 2 and 3 as discussed in the text. Inset shows the temperature
variation of dc magnetic susceptibility.}
\end{figure}

\begin{figure}
%\centerline{\psfig{figure=finepr3.eps,width=15cm,height=17cm}}
\label{Fig 3}
\caption{
Variation of asymmetry parameter $\alpha$ as a function of temperature
for single crystal data.
The inset shows the temperature dependence of resistivity (taken from Ref.
22).
 }
\end{figure}

\begin{figure}
%\centerline{\psfig{figure=finepr4.eps,width=15cm,height=17cm}}
\label{Fig 4}
\caption{Lineshape parameters for powder sample - linewidth, g and
intensity as a function of temperature obtained by fitting spectra to
a Lorentzian profile.
}
\end{figure}

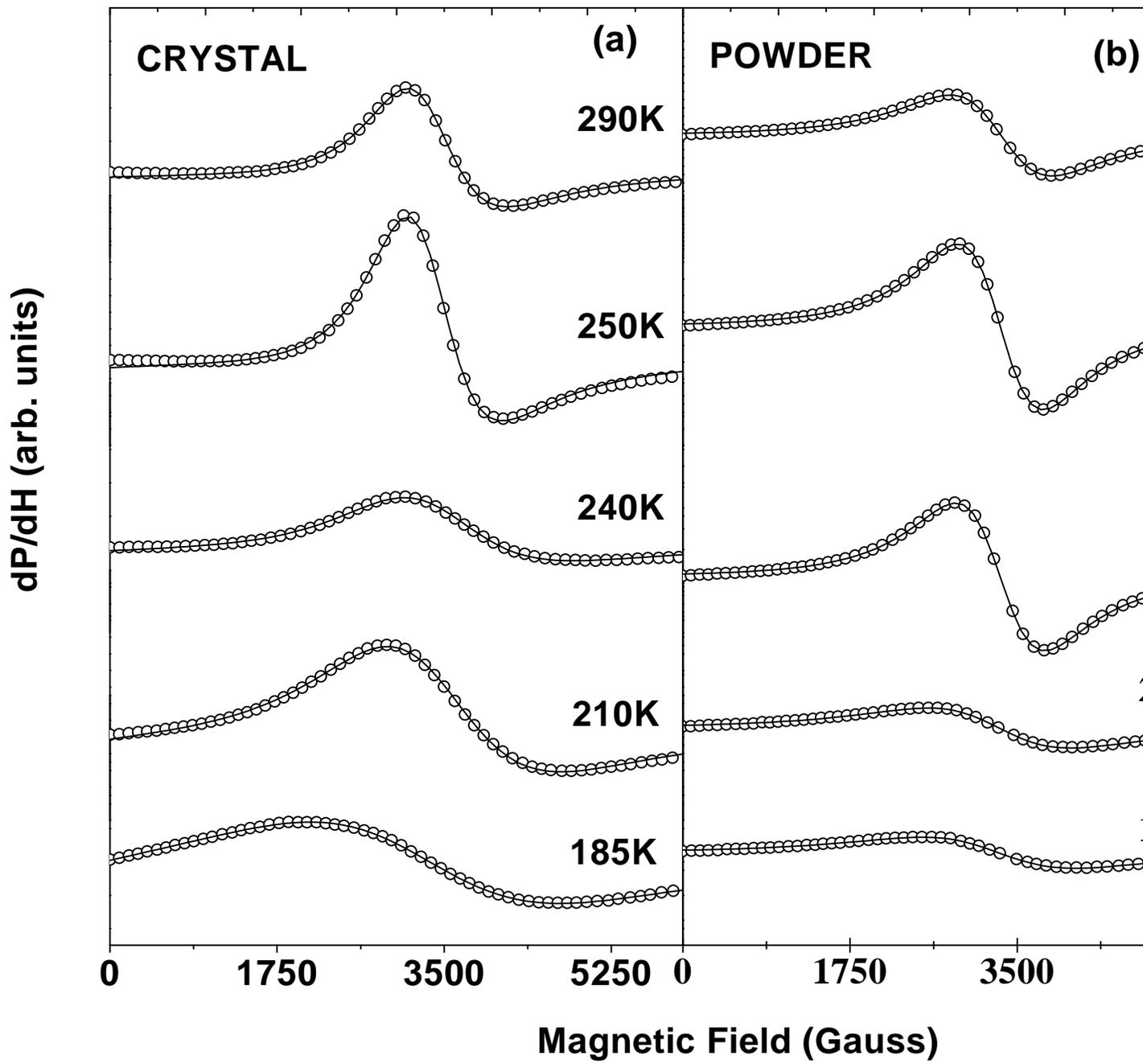
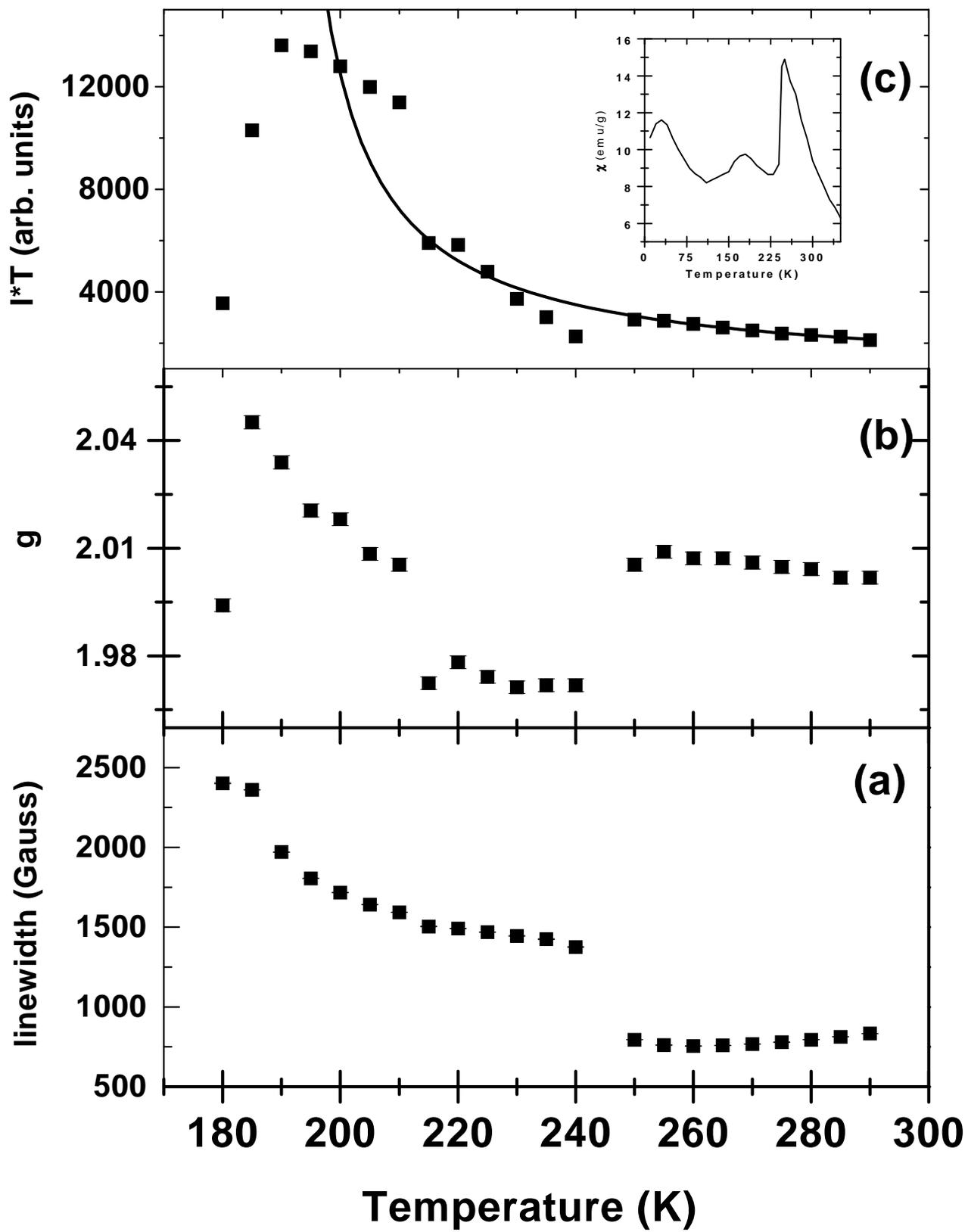
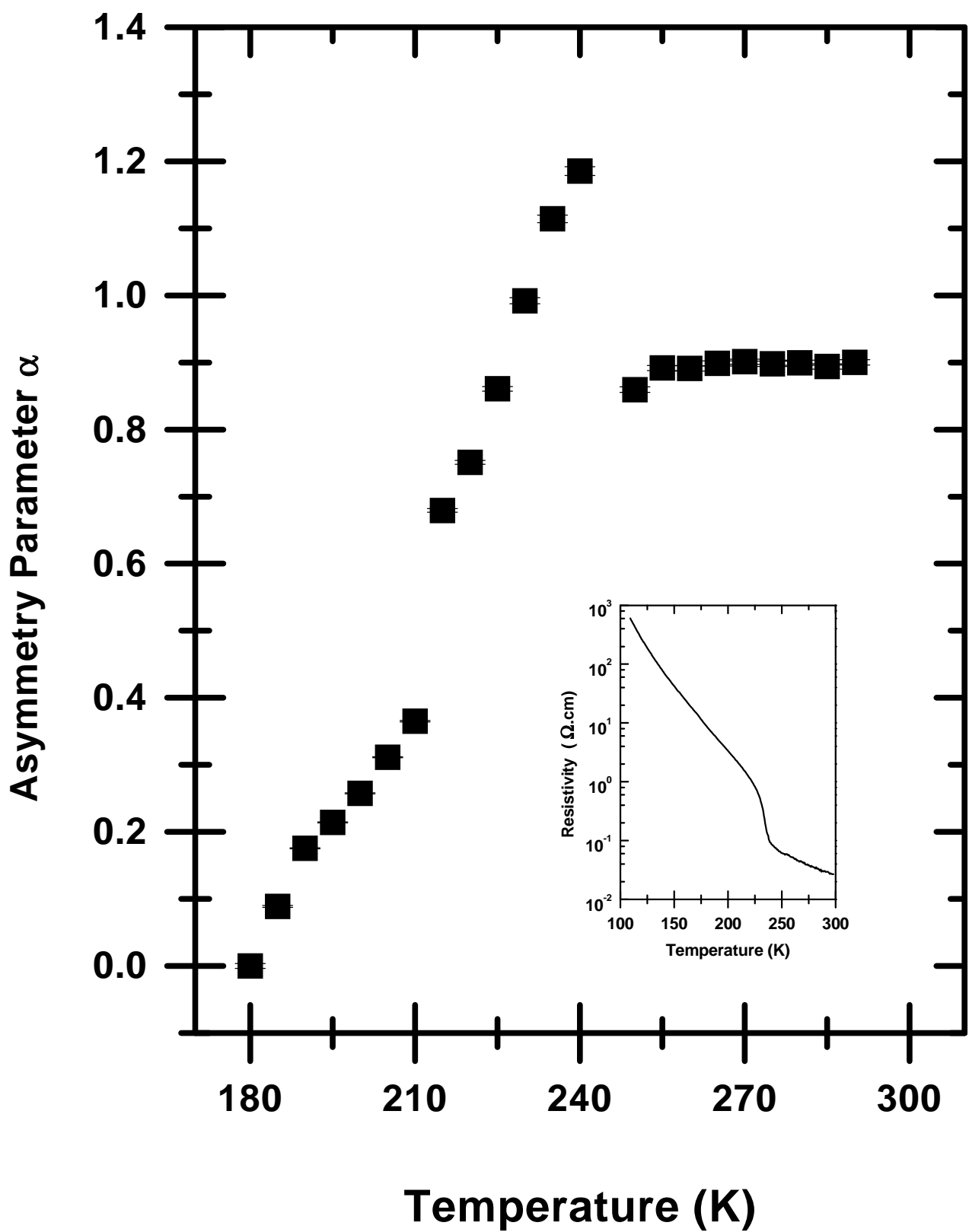
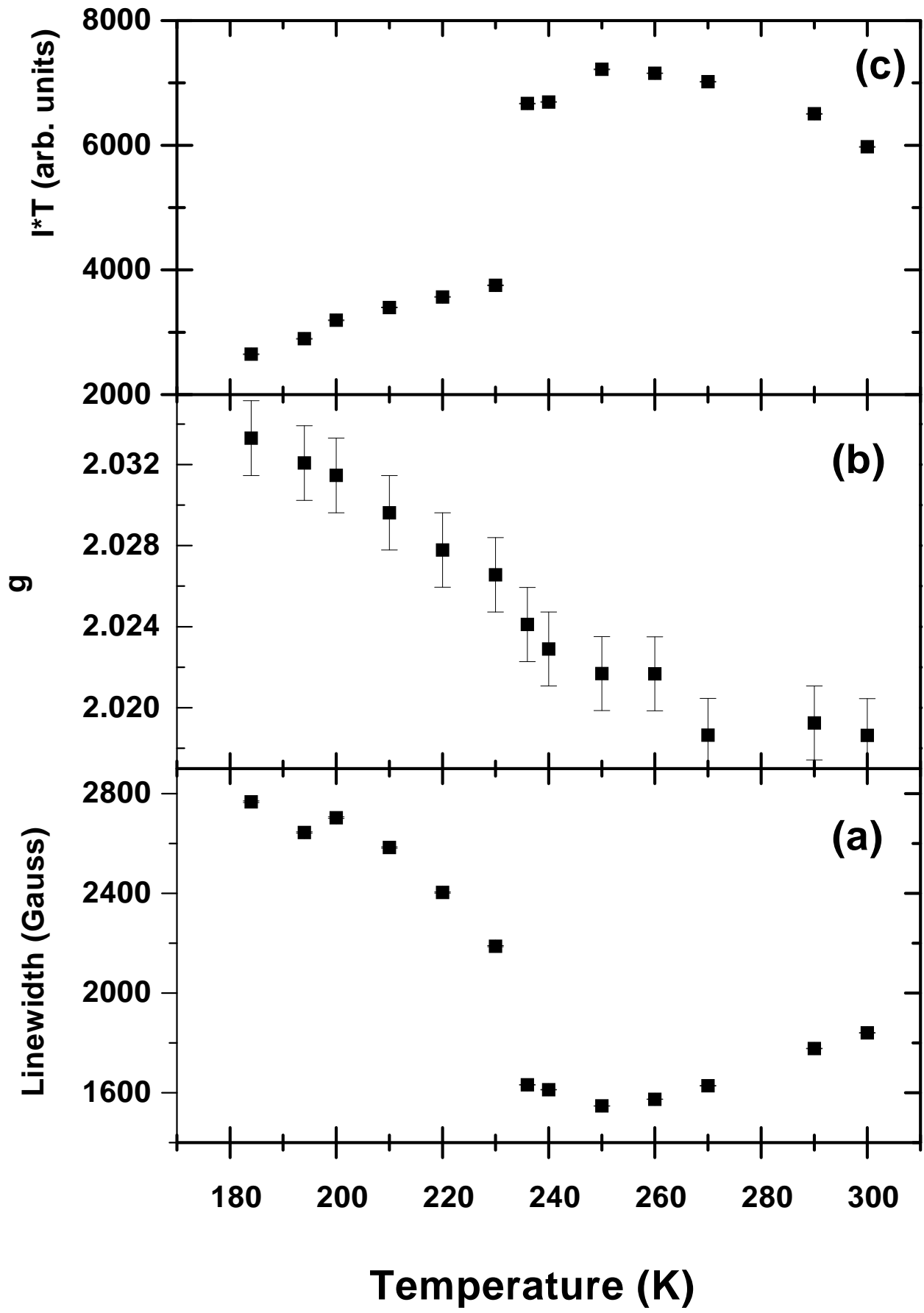
\end{document}